\documentclass[aps,prl,preprint]{revtex4}
\usepackage{graphicx}
\usepackage{appendix}
\usepackage{bm}
\usepackage{subfigure}
\usepackage{amsmath,amsfonts,amssymb}
\DeclareGraphicsExtensions{.eps}
\begin{document}
%%%%%%%%%%%%%%%%%%%%%%%%%%%%%%%%%%%%%%%%%%%%%%%%%%%%%%%%%%%%%%%%%%%%%%%%%%%%
\title{Coefficient of Restitution for Viscoelastic Materials}
%%%%%%%%%%%%%%%%%%%%%%%%%%%%%%%%%%%%%%%%%%%%%%%%%%%%%%%%%%%%%%%%%%%%%%%%%%%%
%
\author{Sangrak Kim}
\email[E-mail address: ]{srkim@kgu.ac.kr}
\affiliation{Department of Physics,\\ Kyonggi University,\\ 94-6 Eui-dong, Youngtong-ku, Suwon 440-760, Korea}
\date{\today}
%
%%%%%%%%%%%%%%%%
\begin{abstract}
%%%%%%%%%%%%%%%%
An analytical expression of the coefficient of restitution for viscoelastic materials
is derived for the viscous-dominant case, such as collisions of polymeric melt. The
recently proposed normal impact force model between two colliding viscoelastic
droplets is employed. To analytically solve the nonlinear differential equation of
deformation caused by the other droplet, a perturbation method is applied. By
combining the forward collision and the inverse collision, we can get an
analytical expression for the coefficient of restitution, which can also be
used to further analyze other properties related with the viscoelasticity.

%%%%%%%%%%%%%%%%
\end{abstract}
%%%%%%%%%%%%%%%%
%

\maketitle
%
%%%%%%%%%%%%%%%%%%%%%%%
\section{introduction}
%%%%%%%%%%%%%%%%%%%%%%%
A coefficient of restitution of an object is defined as a ratio,
\begin{equation}
    \varepsilon = -\frac{\tilde{g}}{g}, \label{def:cof}
\end{equation}
where $g$ and $\tilde{g}$ are normal components of the relative velocities of the colliding objects before collision and after collision, respectively. It is one of the fundamental quantities in characterizing the material property of the granular object. The behavior of granular materials has been of great scientific and technological interests in recent years \cite{schwager98} - \cite{brill2}. Actually, inelastic collisions between granular materials are very common in nature and industry, for example, avalanches of snow or land, flow of sands, powders or cereals and furthermore, astronomical objects such as planetary ring, or stellar dusts.

The normal impact force model between two viscoelastic bodies has recently been proposed \cite{else}-\cite{jkps}. The normal contact force between two colliding bodies can be decomposed into an elastic part and a dissipative part. The elastic component $F_{el}$ is given as,
\begin{equation}
   F_{el}(\xi) ~=~ m_{eff} D_1 \xi^{\alpha}, \label{eq:genforceA}
\end{equation}
where $\xi$ is a deformation and $m_{eff}$, their effective mass and $D_1$ is a constant of the material. The dissipative component $F_{dis}$ is given as,
\begin{equation}
F_{dis}(\xi, \dot{\xi}) ~=~ m_{eff} D_2 \xi^{\gamma} \dot{\xi}, \label{eq:genforceB}
\end{equation}
where $D_2$ is also a constant of the material.

The viscoelastic materials can be classified into two categories: viscous-dominant and elastic-dominant. For this classification, a ratio $D_* \equiv {D_1}/{D_2}^{\frac{\alpha-1}{\gamma}}$ is defined. If $D_* < 1$, it is called a viscous-dominant case; otherwise, it is called an elastic-dominant case. Hard solid objects such as steel balls exhibit the elastic-dominant behavior. On the other hand, soft polymeric objects show the viscous-dominant behavior. Some expressions of the coefficient of restitution for granular materials have already been addressed for the  elastic-dominant cases \cite{schwager98}-\cite{brill2}. In this study, we will derive an expression for the coefficient of restitution for the viscous-dominant case, which is the first attempt. Since we cannot exactly solve with any analytical methods, due to the involved nonlinearity, we have to resort to an approximate perturbation method applied to the corresponding nonlinear differential equation.

In section \ref{sec:deformation}, an equation of the deformation to be solved is introduced. To make the equation into a compact form, we change variables and consequently, we get a single perturbation parameter $\Omega$, defined later. The exact solution for the limit case of $D_*=0$ is easily obtained. In section \ref{sec:forward}, the perturbation solution for the forward collision is addressed. In section \ref{sec:inverse}, the perturbation solution for the inverse collision is presented. To get the proper solution, we have to consider the inverse collision as well as the forward collision process. In section \ref{sec:connection}, a connection between forward and inverse collisions is explained and finally we get an approximate, but analytical expression for the coefficient of restitution.

%%%%%%%%%%%%%%%%%%%%%%%
\section{equation of deformation}
\label{sec:deformation}
%%%%%%%%%%%%%%%%%%%%%%%

The equation of deformation $\xi (t)$ due to the collision is written as
\begin{equation}
\label{def:geneqmotion}
   \ddot{\xi} + D_1 \xi^{\alpha}+D_2 \xi^{\gamma} \dot{\xi} = 0,
\end{equation}
with initial conditions $\xi (0) = 0$ and  $\dot{\xi} (0) = g = {[\vec{v}_1 (0)- \vec{v}_2 (0)]\cdot[\vec{r}_1 (0)- \vec{r}_2 (0)]}/|\vec{r}_1 (0)- \vec{r}_2 (0)|$. Since Eq. (\ref{def:geneqmotion}) has intrinsic nonlinear terms, in general, it cannot be exactly solved. Thus, we have to resort to other approximate methods such as a perturbation solution or numerical integration. Numerical integration results has already been done in Ref. \cite{else}. Here, we will focus on the analytical derivation using the perturbation method.

From now on, we consider the viscous-dominant case $D_*< 1$. First, let us consider extreme case of $D_1 = 0$, \textit{i.e.} with no elasticity at all (damped-only), then Eq. (\ref{def:geneqmotion}) reduces to $\ddot{\xi_0} + D_2 \xi_0^{\gamma} \dot{\xi_0}= 0$, which has an exact solution when $\gamma = 1$,
\begin{equation}
\label{def:exactsol1}
\xi_0(t) = \sqrt{\frac{2g}{D_2}} \tanh(\sqrt{\frac{gD_2}{2}} t).
\end{equation}
In general, $D_1 \neq 0$, so a perturbation method should be employed to get an approximate analytic solution.

To make Eq. (\ref{def:geneqmotion}) into a more compact form, let us change variables as follows:
\begin{subequations}
\begin{align}
\label{def:changed}
   v & \equiv D_2 ^{\frac{1}{\gamma}} g, \\
   \tau & \equiv (D_2 g^{\gamma})^{\frac{1}{\gamma +1}} t =  v^{\frac{\gamma}{\gamma +1}} t, \\
   x & \equiv {(\frac{D_2}{g})}^{\frac{1}{1+\gamma}} \xi = \frac{D_2 ^{\frac{1}{\gamma}}}{v^{\frac{1}{\gamma +1}}} \xi.
\end{align}
\end{subequations}
Then, Eq. (\ref{def:geneqmotion}) is further simplified,
\begin{equation}
\label{eq:scaled1}
   \ddot{x} + \Omega x^{\alpha}  +  x^{\gamma} \dot{x} = 0,
\end{equation}
where $\Omega \equiv {D_\ast} v^{{\alpha -1 -2 \gamma}/{\gamma +1}} $, $\dot{x} \equiv {dx}/{d\tau}$ and $\ddot{x} \equiv {d^2 x}/{d\tau ^2}$, and the initial conditions are accordingly changed as $x(0) =0, ~~\dot{x}(0) = 1 $. Eq. (\ref{eq:scaled1}) is characterized only by one parameter $\Omega$, instead of two parameters $D_1$ and $D_2$ as in Eq. (\ref{def:geneqmotion}). Thus, $\Omega$ can be taken as a perturbation parameter. Note that Eq. (\ref{eq:scaled1}) seems to be independent of $g$.

Again, if $\Omega = 0$, then we get a damped-only equation,
\begin{equation}
\label{eq:dampsol}
   \ddot{x_0} + x_0^{\gamma} \dot{x_0} = 0.
\end{equation}
When $\gamma = 1$, this can also be solved analytically to give as,
\begin{equation}
\label{eq:onlysol}
   x_0(\tau) = \sqrt{2 } \tanh(\frac{\tau}{\sqrt{2 }}),
\end{equation}
which is actually the same as Eq. (\ref{def:exactsol1}). Taylor expansion of Eq. (\ref{eq:onlysol}) is given as,
\begin{equation}
\label{eq:taylor}
   x_0(\tau) = \tau - \frac{1}{6} \tau^3 + \frac{1}{30} \tau^5 - \frac{17}{2520} \tau^7 + \frac{31}{22680} \tau^9  - \frac{691}{2494800} \tau^{11} +\cdots.
\end{equation}
This is only valid for $\tau < \pi/\sqrt{2}$.

%%%%%%%%%%%%%%%%%%%%%%%
\section{perturbation solution: forward collision}
\label{sec:forward}
%%%%%%%%%%%%%%%%%%%%%%%

For $\Omega \neq 0$, we choose $\eta$ such that
\begin{equation}
\label{eq:xtau1}
   x(\tau) \equiv {\tau}\{1-\eta(\tau)\}.
\end{equation}
Inserting Eq. (\ref{eq:xtau1}) into Eq. (\ref{eq:scaled1}), we get,
\begin{equation}
\begin{split}
\label{eq:scaled2}
    -\tau \ddot{\eta} - 2 \dot{\eta} + {\Omega} \tau^{\alpha} ( 1-\eta )^{\alpha} + \tau ^{\gamma} ( 1-\eta )^{\gamma} (1-\eta -\tau \dot{\eta})= 0.
\end{split}
\end{equation}

If we approximate $\alpha \approx \frac{q_{\alpha}}{p}$, and $\gamma\approx \frac{q_{\gamma}}{p}$ to express them in rational numbers with the least common denominator $p$, where $q_{\alpha}$ and $q_{\gamma}$ are integer values, then we can expand $\eta(\tau)$ in powers of $\tau^{1/p}$ as,
\begin{equation}
\label{eq:expand1}
   \eta(\tau) = \sum_{k=0}^{\infty} a_k {(\tau^{\frac{1}{p}})}^k.
\end{equation}
From the simulation results \cite{else}, $p = 2$, $q_{\alpha} = 3$ and $q_{\gamma} = 2$. Eq. (\ref{eq:scaled2}) can be rewritten as,

\begin{equation}
\label{eq:solved1}
   \ddot{x} + \Omega x^{3/2}  +  x \dot{x} = 0.
\end{equation}

Now suppose that $x(\tau, \Omega)$ is the solution of Eq. (\ref{eq:solved1}) in which the equation and the initial condition depend smoothly on a parameter $\Omega$. We want to know the perturbation solution of the form in Taylor expansion of $\Omega$,
\begin{equation}
\label{eq:taylor}
   x(\tau, \Omega) = x_0 (\tau) + \Omega x_1 (\tau) + \frac{\Omega^2}{2!} x_2 (\tau) + \frac{\Omega^3}{3!} x_3 (\tau) + \cdots,
\end{equation}
where $x_0 (\tau)$ is the solution of unperturbed solution, given by
\begin{equation}
\label{eq:unperturbed}
   x_0(\tau) = \sqrt{2 } \tanh(\frac{\tau}{\sqrt{2 }}).
\end{equation}

If we differentiate Eq. (\ref{eq:solved1}) and the initial conditions with respect to $\Omega$, we get
\begin{equation}
\label{eq:1stsol}
   \ddot{z} + x_{0}^{3/2} + z \dot{x_{0}} + x_{0} \dot{z} = 0,
\end{equation}
where $z=z(\tau)=\frac{\partial x(\tau, \Omega)}{\partial \Omega}|_{\Omega=0}$. This generate initial value problems for $\frac{\partial x}{\partial \Omega}$ for all $\tau$ and $\Omega$. Then we have $x_1 (\tau)= \frac{\partial x(\tau, \Omega)}{\partial \Omega}|_{\Omega=0}$. In the first-order, $x_0 (\tau) + \Omega x_1 (\tau)$ is the desired approximate solution. $z(\tau)$ is given by,
\begin{equation}
\begin{split}
\label{eq:1stsol}
   z(\tau) = c_2 sech^2 \frac{\tau}{\sqrt{2}} + [\int_0 ^\tau \cosh^2 \frac{y}{\sqrt{2}} + {\arctan\sqrt{\tanh} \frac{y}{\sqrt{2}} +c_1 + 2^{1/4} \log(-1+\sqrt{\tanh \frac{y}{\sqrt{2}}})- 2^{1/4} \log(1+\sqrt{\tanh \frac{y}{\sqrt{2}}}+4 2^{1/4} \sqrt{\tanh \frac{y}{\sqrt{2}}}}dy] sech^2 \frac{\tau}{\sqrt{2}}.
\end{split}
\end{equation}

Since $\eta(0) = 0$, we get $a_0 =0$. Furthermore, for $\eta(\tau)$ to have a finite value at $\tau = 0$, we also have $a_1 =0, a_2 =0, \cdots, a_{p-1} =0$ for $p>1$.

Inserting Eq. (\ref{eq:expand1}) into Eq. (\ref{eq:scaled2}) and comparing term by term in ${(\tau^{\frac{1}{p}})}^k$, the coefficients $a_k$ can be evaluated, then we have finally,
\begin{equation}
\begin{split}
\label{eq:x}
   x(\tau) = x_0 (\tau) +  {\Omega} x_1 (\tau) + {\Omega}^2 x_2 (\tau)
   + {\Omega}^3 x_3 (\tau) + {\Omega}^4 x_4 (\tau) + \cdots,
\end{split}
\end{equation}
where
\begin{subequations}
\begin{align}
\label{xs}
   x_0 (\tau) & \equiv {\tau} -\frac{1}{6} {\tau}^{3}+ \frac{1}{30} {\tau}^{5}-\frac{17}{2520} {\tau}^{7} +\frac{31}{22680} {\tau}^{9} - \frac{691}{2494800} {\tau}^{11} +\cdots, \\
   x_1 (\tau) & \equiv -\frac{4}{35} {\tau}^{\frac{7}{2}} +\frac{107}{3465} {\tau}^{\frac{11}{2}} -\frac{42683}{5405400} {\tau}^{\frac{15}{2}}  +\frac{2719883}{1396755360} {\tau}^{19/2} + \cdots, \\
   x_2 (\tau) & \equiv \frac{1}{175} {\tau}^{6} - \frac{169}{64680} {\tau}^{8}+\frac{206383}{227026800} {\tau}^{10}+ \cdots, \\
   x_3 (\tau) & \equiv -\frac{22}{104125} {\tau}^{\frac{17}{2}} +\frac{165853}{1096811100} {\tau}^{21/2} + \cdots, \\
   x_4 (\tau) & \equiv \frac{52}{8017625} {\tau}^{11} + \cdots.
\end{align}
\end{subequations}
Note that the first term $x_0 (\tau)$ in Eq. (\ref{xs}) is the same as Eq. (\ref{eq:taylor}), the perturbation solution of the damped-only collision with $\Omega = 0$.

\begin{figure}[htbp]
\begin{center}
\includegraphics[scale=.6]{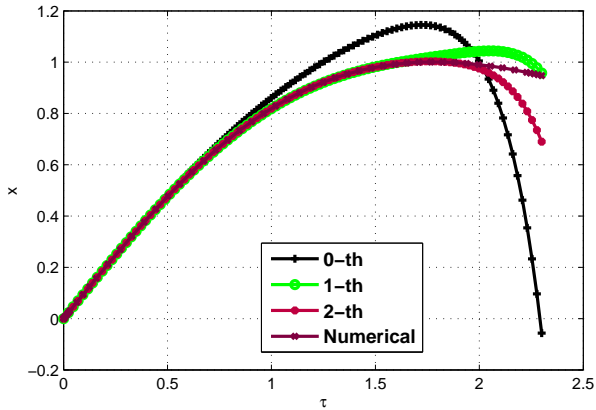}
\caption{(Color on-line) Numerical solution and perturbed solutions for forward collision with an impact speed $g = 0.1$.}
\label{fwdpertsols}
\end{center}
\end{figure}

The time dependency of $x(\tau)$ of the first two successive orders of perturbation solutions is shown in Fig. \ref{fwdpertsols} for an impact velocity of $g = 0.1$. The numerical solution is also shown for a reference. Note that in general, there is no restriction for $\tau$ values, but only when $\tau < 1$, the series Eq. (\ref{eq:expand1}) may absolutely converge. The perturbation solutions higher than the first order begin to diverge around $\tau \approx 2.0$, which is the same magnitude of the bound time $\tau_b$, at which point the forward solution will approach a maximum and the velocity of droplets changes sign. From simulation results \cite{else}, $\tau_b = 2.2$ for $g = 0.1$. Note that the first order solution is better than a second or higher order solutions for $\tau >2$. In other words, a second or higher order perturbation solutions become to diverge for $\tau > 2$. Thus, we need another instrument to extend further beyond time $\tau = 2$.

%%%%%%%%%%%%%%%%%%%%%%%
\section{perturbation solution: inverse collision}
\label{sec:inverse}
%%%%%%%%%%%%%%%%%%%%%%%

For the forward collision, the contact starts with a relative speed $g$ and finally at time $\tau = \tau_c$ with a relative speed $\tilde{g}$. We can define an `inverse' collision as a collision that starts at time $\tau_c $ with a relative speed $\tilde{g}$ and moves backwards in time to finish at time $0$ with a relative speed $g$. Here, the duration of the collision $\tau_c $ is given by the condition that $\ddot{x}(\tau_c )=0$ with $\tau_c >0$. The corresponding equation for the inverse collision $\tilde{x}$ is thus given as,
\begin{equation}
\label{eq:phi}
   \ddot{\tilde{x}} + {\tilde{\Omega}}  {\tilde{x}^{3/2}} - {\tilde{x}} \dot{\tilde{x}} = 0,
\end{equation}
where $\tilde{\Omega} \equiv D_\ast \tilde{v}^{\alpha-1-2\gamma)/(\gamma+1)}$, $\tilde{v} \equiv D_2 ^{1/\gamma} \tilde{g}$. Initial conditions at $\tilde{\tau}=0$ should change accordingly as ${\tilde{x}}(0) = \tilde{x}_c$ and $\dot{\tilde{x}}(0) = 1$. Time $\tilde{\tau}$ is measured from $\tau_c$ and decreases towards the time origin, \textit{i.e.} $\tilde{\tau} \equiv \tau_c -\tau$.

If $\tilde{\Omega} = 0$, then we have
\begin{equation}
\label{eq:xin0}
 \ddot{\tilde{x}} - {\tilde{x}} \dot{\tilde{x}} = 0,
\end{equation}
from which we get an analytic solution,
\begin{equation}
\label{eq:xin00}
 \tilde{x}_0 (\tilde{\tau} ) = G \tanh{(\frac{1}{2} \ln{\frac{1+\frac{\tilde{x}_c}{G}}{1-\frac{\tilde{x}_c}{G}} + \frac{1}{2}} G \tilde{\tau})},
\end{equation}
where $G\equiv \sqrt{{\tilde{x}_c}^2 +2}$. In the limit of $\tilde{x}_c \rightarrow 0$, we have simply $\tilde{x}_0 (\tilde{\tau} ) = \sqrt{2} \tan{(\frac{1}{\sqrt{2}} \tilde{\tau})}$, which has a Taylor expansion,
\begin{equation}
\label{eq:taylor1}
   \tilde{x}_0(\tilde{\tau}) = \tilde{\tau} + \frac{1}{6} \tilde{\tau}^3 + \frac{1}{30} \tilde{\tau}^5 + \frac{17}{2520} \tilde{\tau}^7 + \frac{31}{22680} \tilde{\tau}^9  + \frac{691}{2494800} \tilde{\tau}^{11} +\cdots.
\end{equation}
This is only valid for $\tilde{\tau} < \sqrt{2}$.

Following the same procedure as in forward case, we finally get,
\begin{equation}
\begin{split}
\label{eq:x2}
   \tilde{x} (\tilde{\tau}) = \tilde{x}_0 (\tilde{\tau}) +  {\tilde{\Omega}} \tilde{x}_1 (\tilde{\tau}) + {\tilde{\Omega}}^2 \tilde{x}_2 (\tilde{\tau})
   + {\tilde{\Omega}}^3 \tilde{x}_3 (\tilde{\tau}) + {\tilde{\Omega}}^4 \tilde{x}_4 (\tilde{\tau}) + \cdots,
\end{split}
\end{equation}
where
\begin{subequations}
\begin{align}
\label{xsb}
   \tilde{x}_0 (\tilde{\tau}) & \equiv \tilde{\tau} +\frac{1}{6} \tilde{\tau}^{3}+ \frac{1}{30} \tilde{\tau}^{5}+\frac{17}{2520} \tilde{\tau}^{7} +\frac{31}{22680} \tilde{\tau}^{9} + \frac{691}{2494800} \tilde{\tau}^{11} +\cdots, \\
   \tilde{x}_1 (\tilde{\tau}) & \equiv -\frac{4}{35} \tilde{\tau}^{\frac{7}{2}} -\frac{107}{3465} \tilde{\tau}^{\frac{11}{2}} -\frac{42683}{5405400} \tilde{\tau}^{\frac{15}{2}} -\frac{2719883}{1396755360} \tilde{\tau}^{19/2} + \cdots, \\
   \tilde{x}_2 (\tilde{\tau}) & \equiv +\frac{1}{175} \tilde{\tau}^{6} + \frac{169}{64680} \tilde{\tau}^{8} +\frac{206383}{227026800}\tilde{\tau}^{10} + \cdots, \\
   \tilde{x}_3 (\tilde{\tau}) & \equiv -\frac{22}{104125} \tilde{\tau}^{\frac{17}{2}} -\frac{165853}{1096811100}\tilde{\tau}^{21/2} - \cdots, \\
   \tilde{x}_4 (\tilde{\tau}) & \equiv \frac{52}{8017625} \tilde{\tau}^{11} + \cdots.
\end{align}
\end{subequations}

\begin{figure}[htbp]
\begin{center}
\includegraphics[scale=.6]{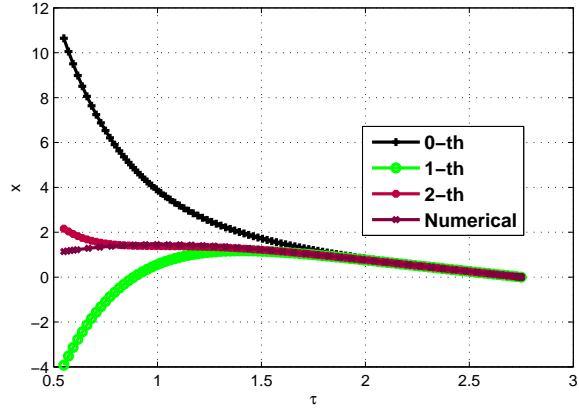}
\caption{(Color on-line) Numerical solution and perturbed solutions for inverse collision with an impact speed $g = 0.1$.}
\label{invpertsols}
\end{center}
\end{figure}

Time dependence of $\tilde{x}(\tilde{\tau})$ of perturbation solutions is shown in Fig. \ref{invpertsols} for the impact velocity $g = 0.1$, where $\tau = \tau_c -\tilde{\tau}$. The numerical solution of Eq. (\ref{eq:phi}) is also shown as a reference. The zeroth-order solution of Eq. (\ref{xsb}
) is identical to the solution for the damped-only case, $\Omega = 0$. We can see that for an inverse collision, second or higher order solutions begin to diverge for $\tau < 1.5$.

\begin{figure}[htbp]
\begin{center}
\includegraphics[scale=.6]{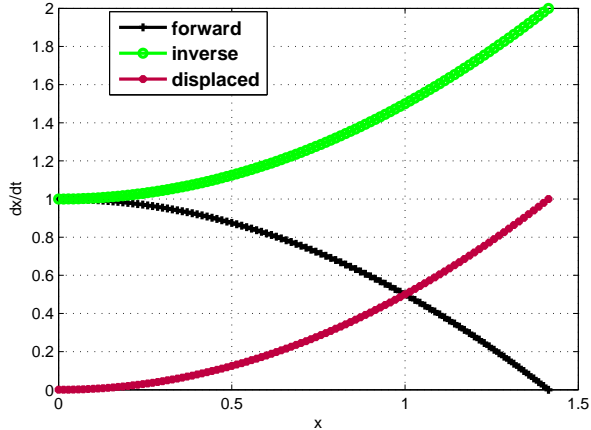}
\caption{(Color on-line) Phase diagrams when $D_1 = 0$. If we displace the whole inverse solution down by -1, then the two solutions cross each other at $x=1$.}
\label{x0phase}
\end{center}
\end{figure}

\begin{figure}[htbp]
\begin{center}
\includegraphics[scale=.6]{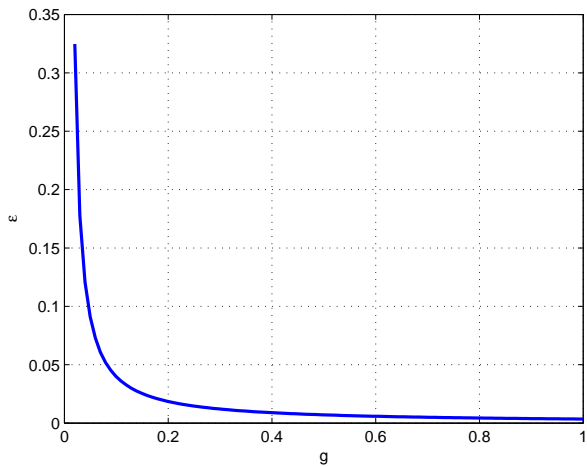}
\caption{(Color on-line) Impact speed $g$ dependence of coefficients of restitution $\varepsilon$ from Eq. (\ref{omegaexpand}).}
\label{conpertsols}
\end{center}
\end{figure}

%%%%%%%%%%%%%%%%%%%%%%%
\section{connection of forward and inverse collision}
\label{sec:connection}
%%%%%%%%%%%%%%%%%%%%%%%

As we pointed out previously, in general, the perturbed solutions diverge around $\tau = \tau_b$. To determine the coefficient of restitution $\varepsilon$, we need the whole time solution from $\tau=0$ to $\tau=\tau_c$. Thus, we have to divide the problem into two parts. The first part is from time $\tau = 0$ to time $\tau_b$ and the second from time $\tau_b$ to time $\tau_c$. We cannot extend the forward collision to the second part, since it diverges from $\tau = 2$, so we have to employ the inverse collision. The perturbation solutions for the forward collision and the inverse collision should be connected smoothly anyway in between. To satisfy this condition, we must have
\begin{subequations}
\begin{align}
   \dot{x} (\tau_b) & = 0, \label{x31} \\
   \dot{\tilde{x}} (\tilde{\tau_b}) & = 0, \label{x32} \\
   x (\tau_b) & = \tilde{x} (\tilde{\tau_b}) \label{x33}.
\end{align}
\end{subequations}

Unfortunately, we cannot find any special points in the both damped-only solutions of Eq. (\ref{eq:onlysol}) and (\ref{eq:xin00}). To use these solutions as a reference, however, we should locate a specific point in the solution as a reference time from which a deviation of the bound time $\tau_b$ can be defined. The first integrals from Eq. (\ref{eq:dampsol}) and (\ref{eq:xin0}) can be used for this purpose,
\begin{subequations}
\begin{align}
\label{xb}
   \dot{x_0}(\tau) + \frac{1}{2}({{x_0}(\tau)})^2 & = 1, \\
   {\dot{\tilde{x}}_0} (\tilde{\tau})- \frac{1}{2}({{\tilde{x}_0}(\tilde{\tau})})^2 & = 1.
\end{align}
\end{subequations}

They are depicted in Fig. \ref{x0phase}. If we displace the whole inverse solution down by -1, then the two solution cross each other at vertical line $x=1$. Let the times at which the forward and inverse solutions meet with the $x=1$ line be $\tau_0$ and $\tilde{\tau}_0$, respectively. These values turn out to be close to $\tau_b$. Thus, this can be used as a reference point for both forward and inverse collisions. Actually, the bound time $\tau_b$ occurs around at $\tau =\tau_0$ where $x(\tau_0) = 1$ and at $\tau = \tilde{\tau}_0$ where $\tilde{x}(\tilde{\tau}_0) = 1$, so it can be set that $\Delta\equiv \tau_b - \tau_0$ and $\tilde{\Delta} \equiv \tilde{\tau}_b - \tilde{\tau}_0$. From Eqs. (\ref{x31}) and (\ref{x32}), we get, to the lowest order, in $\Omega$ and $\tilde{\Omega}$,
\begin{subequations}
\begin{align}
\label{deltas}
  \Delta & = 1+2\dot{x}_1 (\tau_0 ) \Omega, \notag \\
  \tilde{\Delta} & = -1-\frac{2}{3}\dot{\tilde{x_1}} (\tilde{\tau}_0 ) \tilde{\Omega}.
\end{align}
\end{subequations}

Assume that we express $\tilde{\Omega}$ in terms of $\Omega$,
\begin{equation}
\label{omegaexpand}
  \tilde{\Omega} = c_0 + c_1 \Omega + c_2 \Omega^2 +c_3 \Omega^3 + \cdots.
\end{equation}
If we expand Eq. (\ref{x31}) and (\ref{x32}) in terms of $\Delta$ and $\tilde{\Delta}$, respectively, up to second order, then we can determine the coefficients $c_0, c_1, c_2, \cdots$, by comparing term by term. In this way, we can get values for $\varepsilon$ to any accuracy after some algebra. The first three coefficients are calculated to be $c_0 = 6.0203$, $c_1 = -1.5257$, $c_2 = 0.0256$. Finally, the calculated coefficients of restitution are shown in Fig. \ref{conpertsols}. In this graph, we used the material parameters $D_1 = 0.000296$ and $D_2 = 0.0109$ from the simulation results \cite{else}.

%%%%%%%%%%%%%%%%%%%%%%%
\section{Conclusions}
\label{sec:conclusion}
%%%%%%%%%%%%%%%%%%%%%%%

From the previous simulation results \cite{else}, we know that $D_2 = 0.0109$ is much greater than $D_1 = 0.000296$. Thus, we have $D_\ast \approx 3 \times 10^{-3}$ and $\Omega \approx 0.5521$ when $g=0.1$, which means that it works as a good perturbation parameter. Since $\Omega \propto g^{-\frac{3}{4}}$, the coefficient of restitution $\varepsilon$ has terms that are a function of $g^{\frac{1}{4}}$, which can be compared to $g^{\frac{1}{5}}$, as was reported in Brilliantov \textit{et al.} \cite{brillantov}.

Let us now conclude our results. We divided the viscoelastic materials into two classes. One is the elastic-dominant and the other is viscous-dominant. All the previous works have been addressed for the elastic-dominant case only. We analytically derived the expression of the coefficient of restitution for the viscous-dominant case, by the perturbation method. In addition, the validity of this theory can be tested with experimentally with polymer melts or droplets. Aside from calculations of restitution of coefficients, these perturbation solutions can be used in analysis of any other behaviors related with viscoelasticity as well. In other words, the obtained analytic expressions can be used to analyze and predict other polymeric behaviors in the viscous-dominant regime.

%%%%%%%%%%%%%%%%%%%


\begin{references}
%%%%%%%%%%%%%%%%%%%
 \bibitem{schwager98} T. Schwager and T. P\"{o}schel, Phys. Rev. E \textbf{57}, 650 (1998).
 \bibitem{poschel} T. P\"{o}schel, C. Saluena, and T. Schwager, pp. 173-184 in \textit{Continuous and discontinuous modeling of cohesive frictional materials} P. Vermeer, S. Diebels, W. Ehlers, H. Herrmann, S. Luding, and E. Ramm (eds) (Springer, Berlin, 2001).
 \bibitem{brillantov} N. Brilliantov, F. Spahn, J. Hertzsch, and T. P\"{o}schel, Phys. Rev. E \textbf{53}, 5382 (1996).
 \bibitem{ramirez} R. Ram\'{i}rez,  T. P\"{o}schel, N. Brilliantov, and T. Schwager, Phys. Rev. E \textbf{60}, 4465 (1999).
 \bibitem{Schwager2}  T. Schwager and T. P\"{o}schel, Granular Matter \textbf{9}, 465 (2007).
 \bibitem{eurphys} T. Schwager, V. Becker and T. P\"{o}schel, Eur. Phys. J. \textbf{27}, 107 (2008).
 \bibitem{delayed} T. Schwager and T. P\"{o}schel, Phys. Rev. E \textbf{78}, 051304 (2008).
 \bibitem{schwager} T. Schwager, Phys. Rev. E \textbf{75}, 051305 (2007).
 \bibitem{saito} K. Saitoh, A. Bodrova, H. Hayakawa and N. Brilliantov, Phys. Rev. Lett. \textbf{105}, 238001 (2010).
 \bibitem{muller} P. Muller and T. P\"{o}schel, Phys. Rev. E \textbf{84}, 021302 (2011).
 \bibitem{brill2} N. Brilliantov, N. Albers, F. Spahn, and T. P\"{o}schel, Phys. Rev. E \textbf{76}, 051302 (2007).
 \bibitem{else} S. Kim, Phys. Rev. E \textbf{83}, 041302(2011).
 \bibitem{jkps} S. Kim, J. Kor. Phys. Soc. \textbf{56}, 969(2010); S. Kim, J. Kor. Phys. Soc. \textbf{57}, 1339(2010).
\end{references}
\end{document}